\documentclass[universe,article,accept,moreauthors,pdftex]{mdpi} 

\firstpage{1} 
\pubvolume{1}
\issuenum{1}
\articlenumber{0}
\pubyear{2022}
\copyrightyear{2022}
\externaleditor{{Academic Editor: Laura Magrini, Germano Sacco, Lorenzo Spina} 
} 
\datereceived{06 December 2021} 
\dateaccepted{26 January 2022} 
\datepublished{} 
\hreflink{https://doi.org/} 

\Title{Mapping the Galactic Metallicity Gradient with Open Clusters: The State-of-the-Art and Future Challenges}

\TitleCitation{Mapping the Galactic Metallicity Gradient with Open Clusters: The State-of-the-Art and Future Challenges}

\Author{Lorenzo Spina 
 $^{1,}$*
\orcidA{}, Laura Magrini $^{2}$\orcidB{} and Katia Cunha $^{3,4}$\orcidC{}}

\AuthorNames{Lorenzo Spina, Laura Magrini and Katia Cunha}

\AuthorCitation{Spina, L.; Magrini, L.; Cunha, K.}

\address{%
$^{1}$ \quad Instituto Nazionale di Astrofisica --- Padova Observatory, Vicolo dell'Osservatorio 5, 35122 Padova, Italy\\
$^{2}$ \quad Instituto Nazionale di Astrofisica --- Osservatorio Astrofisico di Arcetri, Largo E. Fermi 5, 50125, Firenze, Italy; laura.magrini@inaf.it\\
$^{3}$ \quad Observat\'{o}rio Nacional, Rua General Jos\'{e} Cristino, 77, Rio de Janeiro 20921-400, RJ, 
 Brazil; katia.cunha@noirlab.edu\\
$^{4}$ \quad Steward Observatory, University of Arizona, 933 North Cherry Avenue, Tucson, AZ 85719, 
 USA}

\corres{Correspondence: spina.astro@gmail.com 
}

\abstract{In this paper, we make use of data collected for open cluster members by high-resolution spectroscopic surveys and programmes (i.e., APOGEE, Gaia-ESO, GALAH, OCCASO, and SPA). These data have been homogenised and then analysed as a whole. The resulting catalogue contains [Fe/H] and orbital parameters for 251 Galactic open clusters. The slope of the radial metallicity gradient obtained through 175 open clusters with high-quality metallicity determinations is \mbox{$-$0.064 $\pm$ 0.007 dex kpc$^{-1}$}. The radial metallicity distribution traced by open clusters flattens beyond R$_{\rm Gal}$=12.1 $\pm$ 1.1 kpc. The slope traced by open clusters in the [Fe/H]-L$_{\rm z}$ diagram is \mbox{$-$0.31 $\pm$ 0.02 10$^{3}$ dex km$^{-1}$ kpc$^{-1}$ s}, but it flattens beyond L$_{\rm z}$=2769 $\pm$ 177 km kpc s$^{-1}$. In this paper, we also review some high-priority practical challenges around the study of open clusters that will significantly push our understanding beyond the state-of-the-art. Finally, we compare the shape of the galactic radial metallicity gradient to those of other spiral galaxies.}


\begin{document}

\section{Introduction}
The radial distribution of the metal content in our Galaxy, the so-called radial  metallicity gradient, is an important observational constraint for models that allow us to study the Galactic formation and evolution scenarios e.g., \citep{chiappini97, prantzos00, Minchev14, grisoni18, molla19}. 
Its shape and time evolution provide observational constraints on the disc formation process, on the role of radial flows and stellar migration \citep{schonrich09, spitoni09, spitoni11, spitoni15, Minchev13, kubryk15a, kubryk15b}, and the nature of the infalling gas and outflows \citep{spitoni19, spitoni20, spitoni21}. 
Among the first models able to reproduce the negative abundance gradient observed in the thin disc of the Milky Way, we recall the model of Matteucci $\&$ Francois (1989, \cite{matteucci89}) in which the Galaxy is assumed to form inside-out, i.e., on much shorter timescales in the inner rather than the outer regions, as originally suggested by Larson (1976, \cite{larson76}) and confirmed later by cosmological models; see e.g., \citep{grand17, vincenzo20}.
Generally, the inside-out disc formation mechanism is reproduced by an interplay between the radial variation of the star formation rate (SFR), and an exponential decrease of the gas infalling on the disc e.g., \citep{chiappini97, molla05, Magrini09, palla20}. However, radial flows and stellar migration also have a role in modifying the observed gradients, for instance stellar migration might act in flattening the stellar metallicity gradients e.g., \citep{spitoni15, vincenzo20}. For a general review, we refer to \citet{matteucci21}.  

The Galactic metallicity gradient has be traced with a large variety of objects, from {\sc H~II} regions \citep{balser11, esteban18, arellano20}, to planetary nebulae \citep{maciel07, maciel15, stanghellini10, stanghellini18}, Cepheids \citep{Genovali14, lemasle13, luck18},  low mass \citep{boeche14, huang17, Anders17} and massive stars 
\citep{Daflon04a, braganca19}. Among those objects, open clusters (OCs) are recognised among the best probes of the gradient and its time evolution, since their
ages and chemical composition  can be determined with higher accuracy than for field
stars. 
Open clusters have been used to track the radial metallicity gradient in our Galaxy since the late 1970s \citep{janes79}. 
The results presented in subsequent works tend to converge on a bi-modal gradient, with a change of slope between the Galactocentric radii (R$_{\rm GC}$) 10 and 16 kpc see, e.g., \citep{bragaglia08, sestito08, Magrini09, friel10, Carrera11a, Yong12, Reddy16,  Jacobson16, Casamiquela19, Donor20, Netopil21}. There is a general consensus for a steeper gradient in the inner disc (R$_{\rm gal}\sim$12--13 kpc),
with a much flatter one in the outer regions. 

Thanks to the age range spanned by the OC population (from few Myr to several Gyr, reaching for the oldest clusters 7--8~Gyr), they provide a unique opportunity to study the time evolution of Galactic metallicity gradient.
Several works have investigated it separating the OCs in age bins see, e.g., \citep{carraro98, friel02, Magrini09, Carrera11a, Cunha16, Anders17}, usually finding a flatter gradient for younger clusters see also \citep{Spina14}. 
The differences in the gradient of young and old clusters has stimulated discussion on the role of migration even in objects more massive than stars,  such as clusters e.g., \citep{Anders17, Quillen18a, chen20}. 

The last few years have seen a renewed interest in the study of open clusters, thanks to the launch of the {\em Gaia} satellite \citep{gaia, gaiadr1, gaiadr2, gaiadr3} and numerous spectroscopic surveys, such as {\em Gaia}-ESO \citep{gilmore12}, APOGEE \citep{majewski17} and GALAH \citep{Buder18} which have devoted a considerable amount of observational time to the study of clusters.
In the present work, benefiting from the homogeneous determinations of cluster sample distances and ages obtained through \mbox{{\em Gaia} \citep{Cantat-Gaudin20}}, we use the data from the three main high-resolution spectroscopic surveys to review the Galactic gradient. 
The paper is structured as follows. In Section \ref{surveys}, we give an overview on the main Galactic High-Resolution Spectroscopic Surveys and Programmes that have determined the metal content of open clusters. In Section \ref{dataset}, we homogenise these metallicity values to a common scale and we derive the Galactic orbital parameters of the targeted open clusters. This dataset is then exploited in Section \ref{gradients}, where we outline and discuss the metallicity distribution traced by open clusters across the Galactic disk. Finally, in Section \ref{challenges}, we describe key scientific challenges that should be addressed in order to significantly advance our knowledge.

\section{An Overview on the Galactic High-Resolution Spectroscopic Surveys and Programmes}
\label{surveys}
In this section, we describe the main characteristics of the Galactic High-Resolution Spectroscopic Surveys and Programmes that have determined the metal content of open clusters. These metallicity values are then used in the following sections to describe the radial metallicity distribution traced by open clusters.

\subsection{APOGEE}

The Sloan Digital Sky Survey/APOGEE is a high-resolution (R$\sim$22,000), near infrared spectroscopic (1.51--1.70 $\upmu$m) survey currently operating in both hemispheres, at Apache Point Observatory and Las Campanas Observatory. The APOGEE/DR16 dataset includes about 430,000 stars, collected between August 2011 and August 2018 using the two 300-fiber APOGEE spectrographs \citep{Wilson19}. Given the overall goal to map further distances in the Galaxy, the survey mostly targets evolved stars, but APOGEE has also observed a significant number of dwarf stars, including M dwarfs and young stellar objects \citep{Beaton21,majewski17}.

\textls[-18]{The Open Cluster Chemical Analysis and Mapping (OCCAM) survey \citep{Frinchaboy13} is an ancillary program of APOGEE that aims at producing a comprehensive and uniform data set for open clusters' chemical abundances.
Several previous studies have made use of APOGEE spectra of open cluster members, including} \citet{Cunha16,Donor18,Souto18,Poovelil20,Price-Jones20,Souto21,Souto16,Cunha15}.
The latest contribution from the OCCAM survey \citep{Donor20} presents the analysis of APOGEE DR16 spectra of 128 open clusters, 71 of which are designated to be ``high quality'' based on the appearance of their color-magnitude diagram. They provide radial velocity estimations of the cluster members, as well as detailed abundances for individual elements (e.g., Fe, O, Mg, S, Ca, Mn, Cr, Cu, Na, Al, and K). In the following sections, we adopt the metallicity values from \mbox{\citet{Donor20}.}


\subsection{Gaia-ESO}
The Gaia-ESO survey (GES) is a large public spectroscopic survey carried on with the spectrograph FLAMES \citep{pasquini02} at VLT \citep{gilmore12, randich13} from the end of 2011 to 2018. 
GES has some unique features with respect to the other spectroscopic surveys: it has been performed on a larger telescope, VLT-UT2, an 8-m class telescope, thus reaching fainter and more distant stellar populations; it has observed at two spectral resolutions with UVES (R$\sim$47,000) and with GIRAFFE (R$\sim$20,000); it has covered all stellar populations and all types of stars in the MW, from pre-main sequence stars to old giants, from young clusters in the solar neighbourhood to the halo; the analysis has been performed with a multi-pipeline strategy allowing an in-depth study of the systematic effects affecting spectral analysis; and it has observed a large number of open clusters. 
\textls[-5]{In particular, this aspect is important for the aim of the present review, since GES observed in each cluster large and unbiased samples of stars and the cluster dataset samples the whole age-distance-metallicity parameter space. In addition, GES has included and re-analysed in a homogeneous way a sample of open clusters from the ESO archive, complementing its original sample.  
The final data release ({\sc idr6}) includes 87 clusters. It provides stellar parameters, metallicity, and elemental abundances, radial velocities, and additional products, such as gravity index, chromospheric activity tracers, mass accretion rate diagnostics, and veiling. 
Several works in recent years have been devoted to the study of open clusters observed by the GES, among many we recall the latest ones:} \citet{BertelliMotta18, Magrini18a, Magrini18b, Magrini21, Prisinzano19, Hatzidimitriou19, Casali19, casali20, Baratella20, Randich20, Jackson20, Bonito20, Gutierrez20, Semenova20, Binks20}. 
In the following sections, we adopt the average abundances from {\sc idr6} for the member stars of \mbox{57 clusters} with ages $\ge$120~Myr published in \citet{Magrini21}. For the youngest clusters, we use the metallicity values from \citet{Spina17,Baratella20}.

\subsection{GALAH}

The Galactic Archaeology with HERMES (GALAH) survey \citep{Buder18,DeSilva15,Martell17} acquires data with the 3.9-m Anglo-Australian Telescope at Siding Spring Observatory though the High Efficiency and Resolution Multi-Element Spectrograph (HERMES). The spectrograph disperses the light at $\sim$28,000, which is then captured by four independent cameras and recorded across four non-contiguous channels (4713--4903, 5648--5873, 6478--6737, and 7585--7887 \AA). 

The GALAH’s latest public release, the GALAH+ DR3 catalog \citep{Buder21}, contains data from the main GALAH survey, and it also includes data from ancillary surveys \citep{Stello17,Sharma18}, which use the same spectrograph, observational setup, and data reduction pipeline as the \mbox{GALAH survey.}

Homogenised chemical abundances from the GALAH+ DR3 and APOGEE DR16 catalogs have been used by \citet{Spina21} to derive chemical abundances of 21 elements, from C to Eu, for 134 open clusters, which are publicly available\endnote{\url{https://vizier.u-strasbg.fr/viz-bin/VizieR?-source=J/MNRAS/503/3279} accessed on 6 December 2021.} 
Following the authors' recommendation, we caution against the chemical abundances of open clusters that have been derived though only one stellar member, as their values could not be accurate and the uncertainties underestimated. Therefore, in order to make this distinction clearer to the reader, we call the group of clusters whose metallicity was derived by \citet{Spina21} through only one stellar member as the \textit{silver sample}, while all the other clusters belong to the \textit{gold sample}.

\subsection{OCCASO}
The Open Cluster Chemical Abundances from Spanish Observatories (OCCASO) survey \citep{Casamiquela16} targets several Northern open clusters to obtain accurate radial velocities and chemical abundances for more than 20 chemical species from high-resolution spectra (\mbox{R $\geq$ 62,000}) using the facilities available at Spanish observatories and complementing the Gaia-ESO observations from the South. 
The sample clusters have age $\geq$ 0.3~Gyr and have six or more stars close to the Red Clump region in the Colour-Magnitude diagram. They are selected mainly in the poorly studied regions in terms of R$_{\rm GC}$, [Fe/H], age and height above the Plane and for calibration purposes.   
The results of the OCCASO survey have been presented in \citet{Casamiquela16, Casamiquela17, Casamiquela18, Casamiquela19}. 
In the present work, we have used the metallicity values of 18 clusters from \citet{Casamiquela19}.

\subsection{SPA}
The Stellar Population Astrophysics (SPA) project is an on-going Large Programme carried on the 3.6 m Telescopio Nazionale Galileo (TNG) at the Roque de los Muchachos Observatory (La Palma, Spain). 
It is providing high-resolution optical and near-infrared spectra with GIARPS, a combination of the HARPS-N (R$\sim$110,000)) and GIANO-B (R$\sim$50,000) spectrographs, of approximately 500 stars near to the Sun, covering a wide range of ages and properties see, for a general description, \citep{origlia19}. Many of the SPA targets belong to open clusters for which stellar parameters of member stars, and in some cases a large variety of elemental abundances, have been derived, as the young open clusters ASCC~123 \citep{frasca19}, the Praesepe cluster \citep{dorazi20}, Collinder~350, Gulliver~51, NGC~7044 and Ruprecht~171 \mbox{\citet{casali20}}, and a sample of 16 clusters in \citet{Zhang21} located at Galactocentric distance between $\sim$7.7 and $\sim$10~kpc. We use, for our sample, the results from all the mentioned SPA papers. 

\section{The Homogenised Dataset}
\label{dataset}
Each of the surveys mentioned above have collected and analysed data following their own strategies, and using different instruments, tools, models, and techniques. As a consequence of this heterogeneity, there are systematics affecting their data products, and in particular the chemical abundances. For this reason, a large homogenised data set of open clusters' metallicities derived by different high-resolution surveys and programmes is highly desiderable. 

In this section, we describe how we assemble such a data set and how we derive the properties of open clusters' Galactic orbits. These results are listed in Table~\ref{dataset}. This is a unique table---available at the CDS---which reports for each open cluster the physical (coordinates, parallaxes, proper motions), orbital (Galactic velocities, orbital actions), and chemical (homogenised iron abundances) of the open clusters.

\begin{table}[H]

\caption{physical, orbital and chemical properties of open clusters---full table available online at \mbox{the CDS.}}
\label{dataset}

\begin{adjustwidth}{-\extralength}{0cm}
\begin{tabular*}{\fulllength}{c@{\extracolsep{\fill}}cccccccccc}

\toprule
\textbf{Cluster} & \textbf{X\_XYZ\_low} & \textbf{X\_XYZ\_med} & \textbf{X\_XYZ\_up} & \textbf{Y\_XYZ\_low} & \textbf{Y\_XYZ\_med} & \textbf{Y\_XYZ\_up} &  \boldmath{$\ldots$} \\
 & \textbf{[kpc]} & \textbf{[kpc]} & \textbf{[kpc]} & \textbf{[kpc]} & \textbf{[kpc]} & \textbf{[kpc]} & \boldmath{$\ldots$} \\ \midrule
 Blanco 1 & 8.1352 & 8.1353 & 8.1354 & 0.0117 & 0.0117 & 0.0117 &$\ldots$ \\
 Gulliver 24 & 8.9004 & 8.9070 & 8.9137 & 1.3807 & 1.3934 & 1.4061 &  $\ldots$ \\
 King 1 & 9.1289 & 9.1338 & 9.1386 & 1.6636 & 1.6721 & 1.6805 & $\ldots$ \\
 FSR 0494 & 10.3318 & 10.4072 & 10.4863 & 3.7180 & 3.8480 & 3.9846 & $\ldots$ \\
 FSR 0496 & 8.9418 & 8.9485 & 8.9555 & 1.3116 & 1.3232 & 1.3351 & $\ldots$ \\
$\ldots$ & $\ldots$ & $\ldots$ & $\ldots$ & $\ldots$ &$\ldots$ & $\ldots$  \\
\bottomrule
\end{tabular*}
\end{adjustwidth}
\end{table}

\subsection{Metallicity Homogenisation}
\label{homogenisation}
The metallicity values derived by different surveys and programmes are homogenised over the APOGEE sample. The choice of APOGEE as the standard calibrator is justified by the fact that this survey has observed a large number of Galactic open clusters spanning a wide range of metallicities. Here, we use the OCCAM sample from \citep{Donor20}, who selected only evolved stars (i.e., log~g $\leq$ 3.7 dex) for their analysis of metallicity gradients. Thus, their abundances can be considered as not affected by atomic diffusion \citep{Souto18,BertelliMotta18,Liu19}.

\begin{figure}[H]

 \includegraphics[width=13cm]{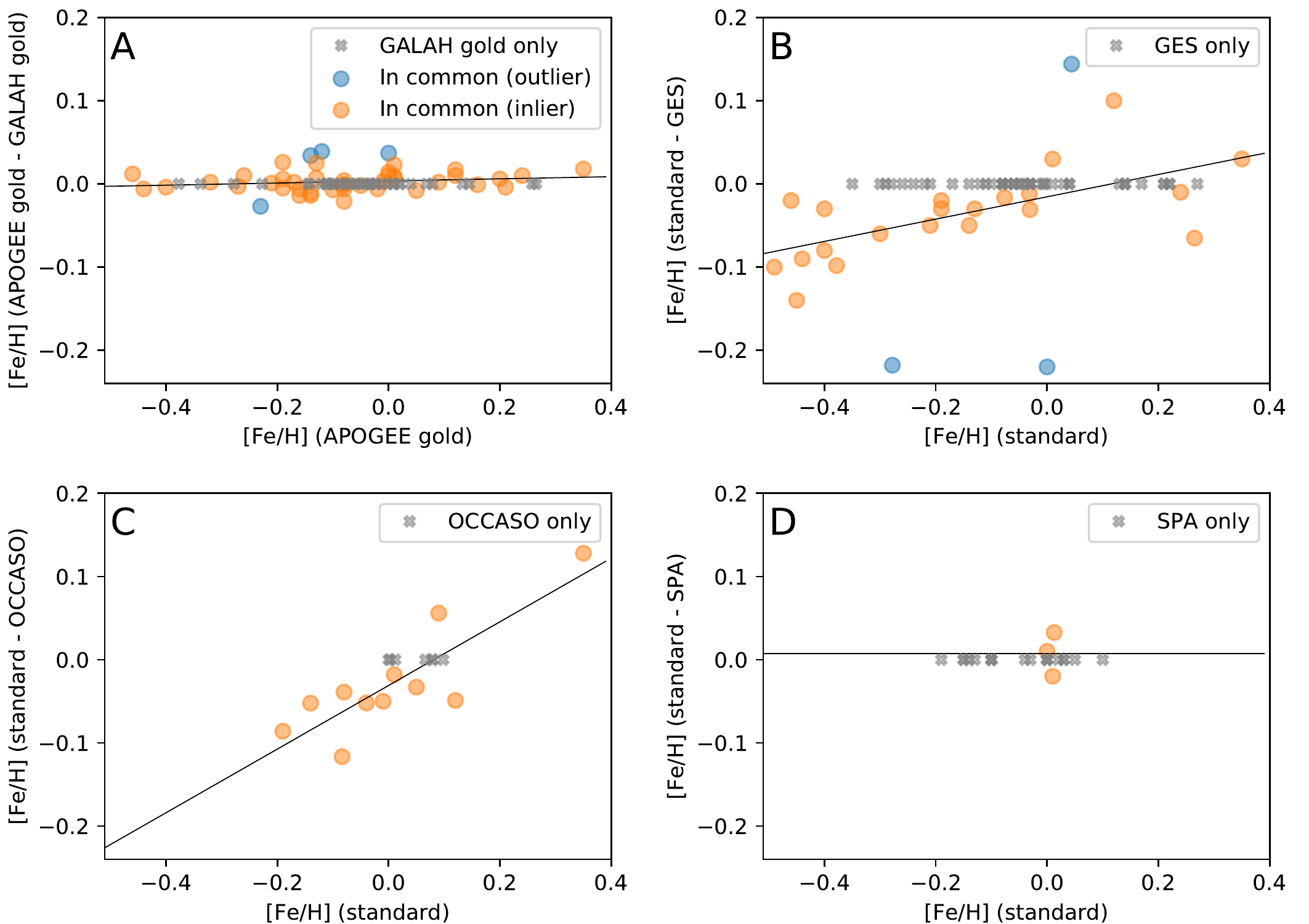}
 \caption{(\textbf{A}--\textbf{D}) The figure shows the different steps of the homogenisation procedure. In each 
 panel, we plot the differential abundances between the clusters in the standard sample and the clusters in the catalog that have to be homogenised, as a function of the [Fe/H] values from the standard. The coloured circles represent the open clusters in common between the two datasets. More specifically, the blue circles are the clusters that are identified as outliers through the Huber regressor, while the orange circles are the inliers. The linear fit is carried out on the latter, and the results are shown as a black line. The grey crosses represent the clusters that are not in common with the standard sample.}
 \label{homo}
\end{figure}

First, we homogenise the GALAH dataset to that of APOGEE. To do so, we use the 44 open clusters in common between APOGEE-gold and GALAH-gold samples. These latter are plotted as coloured (orange or blue) circles in Figure~\ref{homo}A, which shows [Fe/H]$_{\rm APOGEE-gold}$-[Fe/H]$_{\rm GALAH-gold}$ as a function of the [Fe/H] values from the APOGEE-gold sample. The homogenisation is carried out correcting for the linear relation that we assume exists between these two quantities. This relation is found through a Huber regressor \citep{Huber64} that is able to identify the eventual outliers (represented as blue circles in Figure~\ref{homo}) and carry out the linear fitting only of the inliers (orange circles). The hyperparameter \textit{epsilon} of the Huber regressor is set to 2. The result of the linear fit is represented as a black line. Figure~\ref{homo}A also shows as grey crosses the open clusters included in the GALAH-gold sample, but not in the standard sample. These are useful to visually check that the [Fe/H] range of clusters to be homogenised is within the [Fe/H] range of clusters in common with the standard sample. The plot of Figure~\ref{homo}A shows four outliers that are excluded from the linear regression: Basel~11b, Berkeley~71, NGC~1857, and NGC~2304. In fact, these latter have [Fe/H] abundances that are less precise than the majority of the other clusters.

Once the homogenisation of GALAH-gold over APOGEE-gold is accomplished, the open clusters that are part only of the GALAH-gold sample are added to APOGEE-gold. This merged dataset is then used as a new standard for the homogeisation of the Gaia-ESO sample. The analysis described above is repeated, and the results are shown in Figure~\ref{homo}B. In addition, in this case, the are outliers whose [Fe/H] uncertainties are higher that the vast majority of the other clusters: Berkeley~44, Collinder~261, and NGC~2264. Once the Gaia-ESO data are homogenised and added to the standard data set, the resulting catalog is then used as a new standard for the homogenisation of the OCCASO data set (see Figure~\ref{homo}C). 

Then, we perform the homogenisation of the SPA dataset. Since this programme only has three clusters in common with the standard sample, the homogenisation is carried out by subtracting the average of [Fe/H]$_{\rm standard}$-[Fe/H]$_{\rm SPA}$, as it is described in Figure~\ref{homo}D.

Finally, we add the clusters of the APOGEE-silver and GALAH-silver samples that are not included in the catalogs mentioned above. The [Fe/H] from the APOGEE-silver sample remain unchanged. Instead, the GALAH-silver sample is homogenised using the same relation shown in Figure~\ref{homo}A.

The final sample with homogenised iron abundances includes 251 open clusters in total. These are divided in 180 \textit{gold} open clusters, plus 71 other clusters whose abundances are taken from either the APOGEE-silver or GALAH-silver samples. These latter are not considered in the analysis of the Galactic metallicity gradients discussed in Section~\ref{gradients}; nevertheless, for completeness, we report them as well in Table~\ref{dataset}.

Instead of using the largest data set as a standard calibrator, one could make a different choice and homogenize over the highest-quality data set. If we exclude OCCASO and SPA, whose data sets are considerably small compared to those of the other surveys, the highest-quality data set is probably the one from Gaia-ESO, as it is obtained from spectra with the highest resolution. Thus, we verified that, using Gaia-ESO as a standard calibrator, the results discussed in the following sections of this paper do not change.

\subsection{Orbital Parameters}
\label{orbital_parameters}
The kinematic properties of open clusters are derived from their astrometric solutions: right ascension ($\alpha$), declination ($\delta$), parallax ($\varpi$), and proper motions ($\mu_{\alpha}$ and $\mu_{\delta}$). These values are taken from the catalog published by \citet{Cantat-Gaudin20} and are based on Gaia DR2 data of stars with probabilities of being members of the clusters $\geq$ 0.7. Among the 251 open clusters, 216 have astrometric solutions. 
In particular, out of the \mbox{180 \textit{gold}} open clusters, 175 are with astrometric solutions. The remaining five clusters are the star forming regions Carina, Chamaeleon~I, NGC~6530, and Rho Ophiuchi, and the old cluster FSR~0394. When the astrometric solutions are not available from \citet{Cantat-Gaudin20}, we use those from \citet{Kharchenko13}.

Radial velocity (RV) is another fundamental ingredient for the derivation of orbital parameters. These values are taken from the same catalogs listing the iron abundances used in Section~\ref{homogenisation}. In addition to those, we also used the dataset from \citet{Soubiran18} (hereafter, Gaia-RV), which is based on Gaia DR2 radial velocity values.

In order to collect RV determinations for the clusters in our dataset, we give the highest priority to the values based on more than one stellar members and with standard deviation $<$ 3~km~s$^{-1}$ from the GALAH \citep{Spina20}, APOGEE \citep{Donor20}, and Gaia-RV \citep{Soubiran18} datasets, respectively. For the clusters without a RV determination in those catalogues, we use the values from Gaia-ESO \textbf{\citep{gilmore12} and SPA \citep{origlia19}}. Finally, for the remaining clusters, we use the GALAH, APOGEE, and Gaia-RV datasets without restrictions on the number of stellar members and standard deviation. 

The kinematic properties of the open clusters are determined following the same procedure applied to the GALAH open clusters, which is detailed in \citet{Spina20}. First, physical distances from the Sun of every open cluster are obtained from the $\varpi$ value and its uncertainty, processed by \texttt{abj2016}\endnote{Code available at \url{https://github.com/fjaellet/abj2016} accessed on 6 December 2021. Our analysis is carried out using the default settings.}. Following the formalism described in \citet{Astraatmadja16}, this code derives the distance probability density function of the cluster from which we calculate the median value. The latter, ($\alpha$, $\delta$, $\mu_{\alpha}$, $\mu_{\delta}$, and RVs are then transformed into Galactocentric coordinates and velocities, both Cartesian (X, Y, Z, U, V, and W) and cylindrical (R$_{\rm Gal}$, $\phi$, z, v$_{\rm R}$, v$_{\rm \phi}$, v$_{\rm z}$) through \texttt{GALPY} \endnote{Code available at \url{http://github.com/jobovy/galpy} accessed on 6 December 2021.} \citep{Bovy15} a Python package for galactic dynamics. We also compute actions J$_{\rm r}$, L$_{\rm z}$, J$_{\rm z}$, guiding radii r$_{\rm guid}$, eccentricities e, and orbit boundary information (z$_{\rm max}$, R$_{\rm peri}$, and R$_{\rm apo}$) in the Galactic potential \texttt{MWpotential2014} described in \citet{Bovy15} and a Staeckel fudge with 0.45 as the focal length of the confocal coordinate system. 

The statistical uncertainties of all these properties are obtained from a Monte Carlo simulation with a 10,000 sampling size. These samples are randomly drawn from normal distributions centred on $\alpha$, $\delta$, $\mu_{\alpha}$, $\mu_{\delta}$, and RV and with a standard deviation equal to their standard error. For the distance parameter, we use the probability density function obtained through \texttt{abj2016}. In Table~\ref{dataset}, we list the median values and standard deviations of the final distributions of these kinematic properties.

\section{Galactic Metallicity Gradients}
\label{gradients}

Open clusters are tracers of the chemical evolution of the Galactic disk both in space and time. In Figure~\ref{gradients}, we use the information derived in Section~\ref{dataset} to study how metallicity varies across the disk. Here, we consider the \textit{gold} sample of open clusters that are also included in the catalog of \citet{Cantat-Gaudin20}. These 175 open clusters currently represent the largest sample of clusters ever used to trace the Galactic metallicity gradient with high-resolution spectroscopy (i.e., R~$\gtrsim$~20~k). Figure~\ref{gradients}-top shows Fe abundances as a function of the Galactocentric distances in cylindrical coordinates R$_{\rm Gal}$. Clusters are colour coded as a function of their age from \citet{Cantat-Gaudin20}. 

Similarly to many other studies before us \citep{Twarog97,Carrera11a,Yong12,Frinchaboy13,Reddy16,Netopil16,Donor20}, we also model the distribution of open clusters in the [Fe/H]-R$_{\rm Gal}$ space with a broken-line defined as follows: 
\begin{equation}
\label{model}
    y = 
    \begin{cases}
      a_{1} + b_{1}\times x & x \leq k\\
      (b_{1} \times k + a_{1}) + b_{2} \times x & x > k
    \end{cases}       
\end{equation}

Through a Monte Carlo Marcov Chain simulation, we derive the probability density distributions of the parameters in Equation~\eqref{model}. During the simulation, the x$_{i}$ and y$_{i}$ values are randomly drawn from normal distributions centred on the R$_{\rm Gal}$ and [Fe/H] values of the $i{\rm th}$ cluster:
\begin{equation}
\label{eq}
\begin{array}{l}
x_i = \mathcal{N}(R_{\rm Gal, {\it i}}, \delta R_{\rm Gal, {\it i}})\\
y_i = \mathcal{N}([Fe/H]_i, \delta Fe_{i})
\end{array}
\end{equation}
where $\delta$~R$_{{\rm Gal},i}$ is the uncertainty related to R$_{{\rm Gal},i}$ and listed in Table~\ref{dataset}, while $\delta$~Fe$_{i}$ is the quadratic sum between the standard error of [Fe/H] calculated among the members of the $i{\rm th}$ cluster\endnote{Given a standard deviation $\sigma$ measured over a population of size N, the standard error is defined as $\sigma$/$\sqrt{N}$.} and a free parameter $\epsilon$ which accounts for the intrinsic chemical scatter between clusters at fixed Galactocentric radius. In fact, a variety of processes are responsible for this additional scatter in Fe that cannot be explained by measurement uncertainties, such as chemical evolution, radial migration, and the fact that we are projecting a 3D space (e.g., X,Y,Z) to the one-dimensional, variable R$_{\rm Gal}$.

Priors for a$_{1}$, b$_{1}$, b$_{2}$, and k are chosen to be $\mathcal{N}$(0.6, 0.5), $\mathcal{N}$($-$0.07 kpc$^{-1}$, 0.2 kpc$^{-1}$), $\mathcal{N}$(0.0 kpc$^{-1}$, 0.2 kpc$^{-1}$), and $\mathcal{N}$(13 kpc, 2 kpc), respectively. Our prior for the $\epsilon$ parameter is a positive half-Cauchy distribution with $\gamma$ = 1. We run the simulation with 10,000 samples, half of which are used for burn-in, and a No-U-Turn Sampler \citep{Hoffman11}. The script is written in Python using the $\tt{pymc3}$ package \citep{Salvatier16}.

The convergence of the Bayesian inference is checked against the traces of each parameter and their autocorrelation plots. The 68 and 95$\%$ confidence intervals of the models resulting from the posteriors are represented in the top panel of Figure~\ref{gradients} (top) with red shaded areas. There are few clusters that are located out of the shaded areas, such as the metal-rich NGC 6791 (R$_{\rm Gal}$ = 8.0 kpc; [Fe/H] = 0.35 dex) and the
metal-poor NGC 2243 (R$_{\rm Gal}$ = 11.2 kpc; [Fe/H] = ${-}$0.46 dex). These outliers are old clusters (age $>$ 1 Gyr) that have had the time to migrate significantly across the Galactic disk. The posteriors of the parameters from Equation~\eqref{model} are shown in Figure~\ref{posteriors}-left panels. The mean values, standard deviations and 95$\%$ confidence intervals are also listed in Table~\ref{posteriors_tab}-upper panel. 

According to our results, the knee of the Galactic metallicity gradient is located at \mbox{12.1 $\pm$ 1.1 kpc} from the Galactic centre. This is an inner location compared to that found by \citet{Donor20} (i.e., 13.9 kpc). However, our result is in agreement several other works locating the break between 12 and 13 kpc from the Galactic centre
\citep{Carrera11a,Jacobson16,Yong12,Cunha16,Netopil16,Reddy16}. Nevertheless, the break location found by \citet{Donor20} is still within the 95$\%$ confidence interval of \mbox{our solution.}
\begin{figure}[H]

\hspace{-0.27cm} \includegraphics[width=12cm]{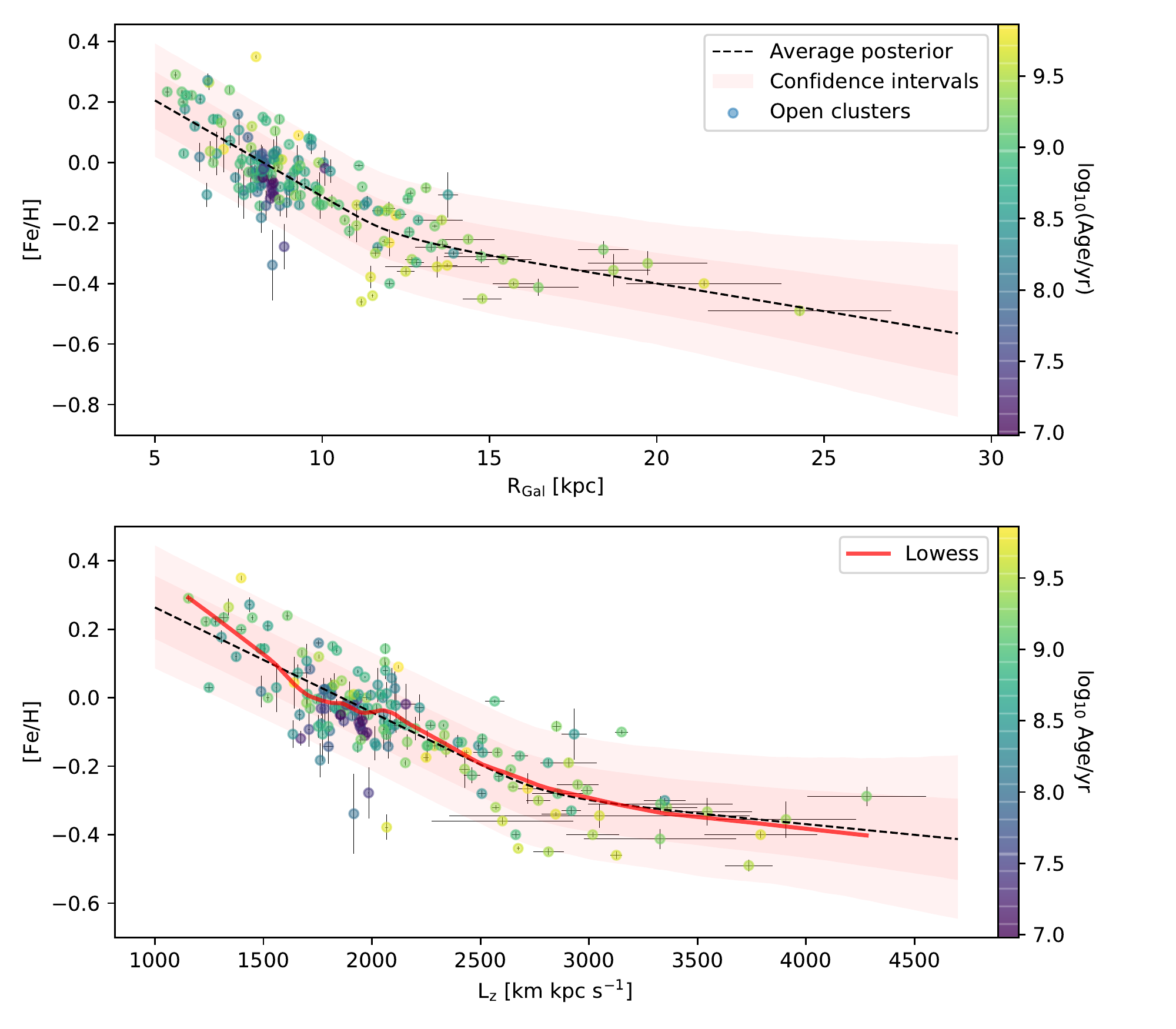}
 \caption{(\textbf{Top}) [Fe/H] values determined for open clusters as a function of their Galactocentric distances R$_{\rm Gal}$. Clusters are colour coded as a function of their  age. Red shaded areas represent the 68 and 95$\%$ confidence intervals of the models resulting from the Bayesian inference, while the black dashed
line traces the most probable model. (\textbf{Bottom}) Same as in the top panel, but with [Fe/H] plotted as a function of angular momentum L$_{z}$.}
 \label{gradients}
\end{figure}

The slope of the inner metallicity gradient (i.e., b$_{1}$) is another important parameter which is tightly linked to the evolution of our Galaxy. Our value of $-$0.064 $\pm$ 0.007 kpc$^{-1}$ is in excellent agreement with many recent works based on open clusters \citep{Netopil16,Carrera19,Casamiquela19,Spina21,Donor20}.

There is a general consensus in the literature that the outer disk has a radial metallicity gradient which is significantly flatter than that of the inner disk. Our results confirm this view. Nevertheless, the slope of the outer gradient traced by our sample \linebreak{(b$_{2}$ = $-$0.019 $\pm$ 0.008 kpc$^{-1}$)} is not consistent with a perfectly flat plateau.

The R$_{\rm Gal}$ values used in Figure \ref{gradients} top panel offer a snapshot of the current location of open clusters, and they are not always representative of where clusters are born. In fact, it is very likely that clusters, especially the oldest ones such as NGC 6791, have migrated a long way from their birth locations. Although with our current information we are not able to trace back all these clusters to their original orbits, the metallicity variation across the Galactic disk can be studied against other parameters that are more fundamental than R$_{\rm Gal}$. For instance, a stellar particle that only interacts with a static axisymmetric potential would see its R$_{\rm Gal}$ changing and following the radial oscillations of its eccentric orbit. On the other hand, its angular momentum L$_{\rm z}$ would stay constant\endnote{However, L$_{\rm z}$ can also change when the particle interacts with non-axisymmetric perturbations of the Galactic potential, such as the bar, spirals and giant molecular clouds.}. Therefore, the main advantage of action integrals---such as L$_{\rm z}$---is that they are conserved along the entire orbit and in fact they are often used to unveil orbital structures in the Galactic disk \citep{Trick19}. In the $pre$-Gaia era, action integrals could only be studied locally ($<\lesssim$500 pc from the Sun) and, for that reason, the metallicity variation across the Galactic disk has been historically studied against R$_{\rm Gal}$. Instead, the precise astrometric solutions now produced by Gaia make it possible to capture the entire kinematic information for a broad range of clusters. This now makes L$_{\rm z}$ a quantity that is much better suited than R$_{\rm Gal}$ to characterise the metal content across the Galaxy \citep{Spina21}. Note that the guiding radius can also be used in place of the angular momentum; in fact, the two are quantities carrying the same information.

\begin{figure}[H]

 \includegraphics[width=14cm]{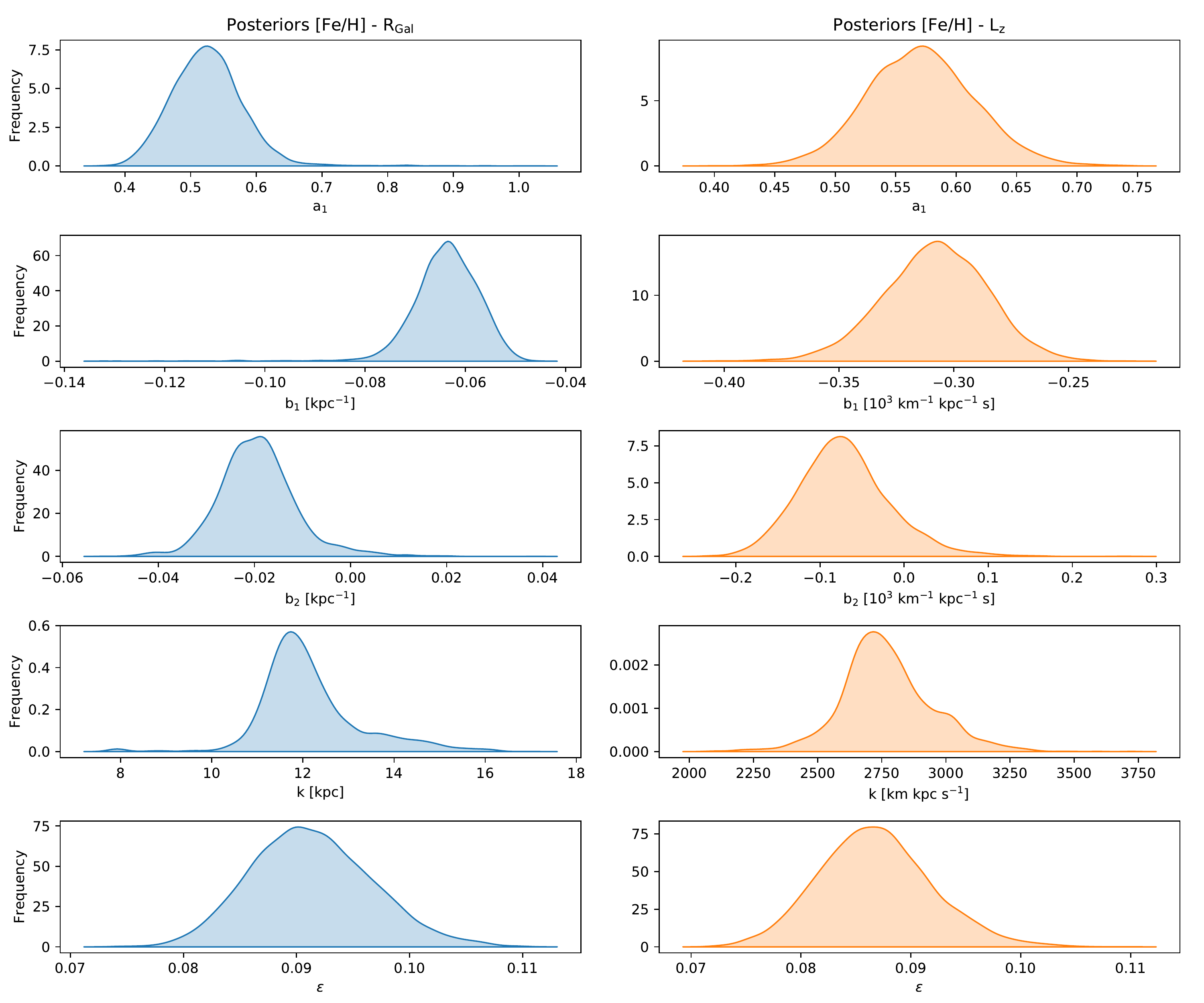}
 \caption{(\textbf{Left}) Posteriors resulting from the modelling in the [Fe/H]-R$_{\rm Gal}$ space (see Figure~\ref{gradients} top). (\textbf{Right}) Posteriors resulting from the modelling in the [Fe/H]-L$_{\rm z}$ space (see Figure~\ref{gradients} bottom). }
 \label{posteriors}
\end{figure}

In Figure \ref{gradients} bottom panel, we show the distribution of open clusters in the [Fe/H]-L$_{\rm z}$ space, which is modelled using Equation~\eqref{model} and the same technique described above. The resulting posteriors are shown in Figure~\ref{posteriors_tab} right panel, and the confidence intervals are also listed in Table~\ref{posteriors_tab} middle panel. 

Similarly to what we observed in the [Fe/H]-R$_{\rm Gal}$ diagram, the distribution of clusters in the [Fe/H]-L$_{\rm z}$ space is also characterised by a steeper inner gradient, a break located around L$_{\rm z}$ $\sim$ 2800 km kpc s$^{-1}$, and a flatter outer slope. Nevertheless, there may be details in the [Fe/H]-L$_{\rm z}$ diagram that our simple model could not capture. For instance, the locally weighted scatterplot smoothing (LOWESS) function in Figure \ref{gradients} bottom shows a hint of a wave-crest between 2000--2200 km kpc s$^{-1}$, which is similar to that found by \mbox{\citet{Wheeler21}} with Gaia and LAMOST data. This particular feature could be linked to the Outer Lindblad Resonance or to the Perseus arm, both located at \mbox{L$_{\rm z}\sim$ 2200 km kpc s$^{-1}$}. As an alternative explanation, this ridge could simply be a statistical fluctuation of the LOWESS function. The urgency of enlarging the current sample of open clusters with metallicity determinations, and the necessity of understanding whether or not the features visible in Figure \ref{gradients} are truly related to resonances of the Galactic disk, are key challenges that lie before us that we will discuss in more details (see Sections~\ref{selection} and \ref{resonances}).

\begin{table}[H]
\caption{Posteriors.}

\label{posteriors_tab}
\begin{tabular*}{\hsize}{c@{\extracolsep{\fill}}ccc}

\toprule
\textbf{Parameter} & \textbf{Mean} & \boldmath{$\sigma$} & \textbf{95\% C.I.}  \\ \midrule
\multicolumn{4}{c}{[Fe/H]---R$_{\rm Gal}$} \\ \midrule
a$_{\rm 1}$ & 0.53 & 0.06 & 0.42--0.64\\
b$_{\rm 1}$ [kpc$^{-1}$] & $-$0.064 & 0.007 & $-$0.076--$-$0.053\\
b$_{\rm 2}$ [kpc$^{-1}$] & $-$0.019 & 0.008 & $-$0.033--$-$0.001\\
k [kpc] & 12.1 & 1.1 & 10.6--14.9\\
$\epsilon$ & 0.091 & 0.005 & 0.082--0.102 \\ \midrule

\multicolumn{4}{c}{[Fe/H]---L$_{\rm z}$} \\ \midrule
a$_{\rm 1}$ & 0.57 & 0.04 & 0.49--0.66\\
b$_{\rm 1}$ [10$^3$ km$^{-1}$ kpc$^{-1}$ s] & $-$0.31 & 0.02 & $-$0.35--$-$0.26\\
b$_{\rm 2}$ [10$^3$ km$^{-1}$ kpc$^{-1}$ s] & $-$0.07 & 0.05 & $-$0.167--$-$0.048\\
k [km kpc s$^{-1}$] & 2769 & 177 & 2429--3156\\
$\epsilon$ & 0.087 & 0.005 & 0.077--0.098 \\ \midrule

\multicolumn{4}{c}{[Fe/H]---R$_{\rm Gal}$, warped disk}  \\ \midrule
a$_{\rm 1}$ & 0.49 & 0.05 & 0.41--0.59\\
b$_{\rm 1}$ [kpc$^{-1}$] & $-$0.060 & 0.005 & $-$0.071--$-$0.050\\
b$_{\rm 2}$ [kpc$^{-1}$] & $-$0.012 & 0.003 & $-$0.019--$-$0.005\\
k [kpc] & 12.3 & 0.5 & 11.1--13.3\\
$\epsilon$ & 0.090 & 0.005 & 0.080--0.101 \\ \bottomrule

\end{tabular*}

\end{table}

\begin{figure}[H]

 \includegraphics[width=13cm]{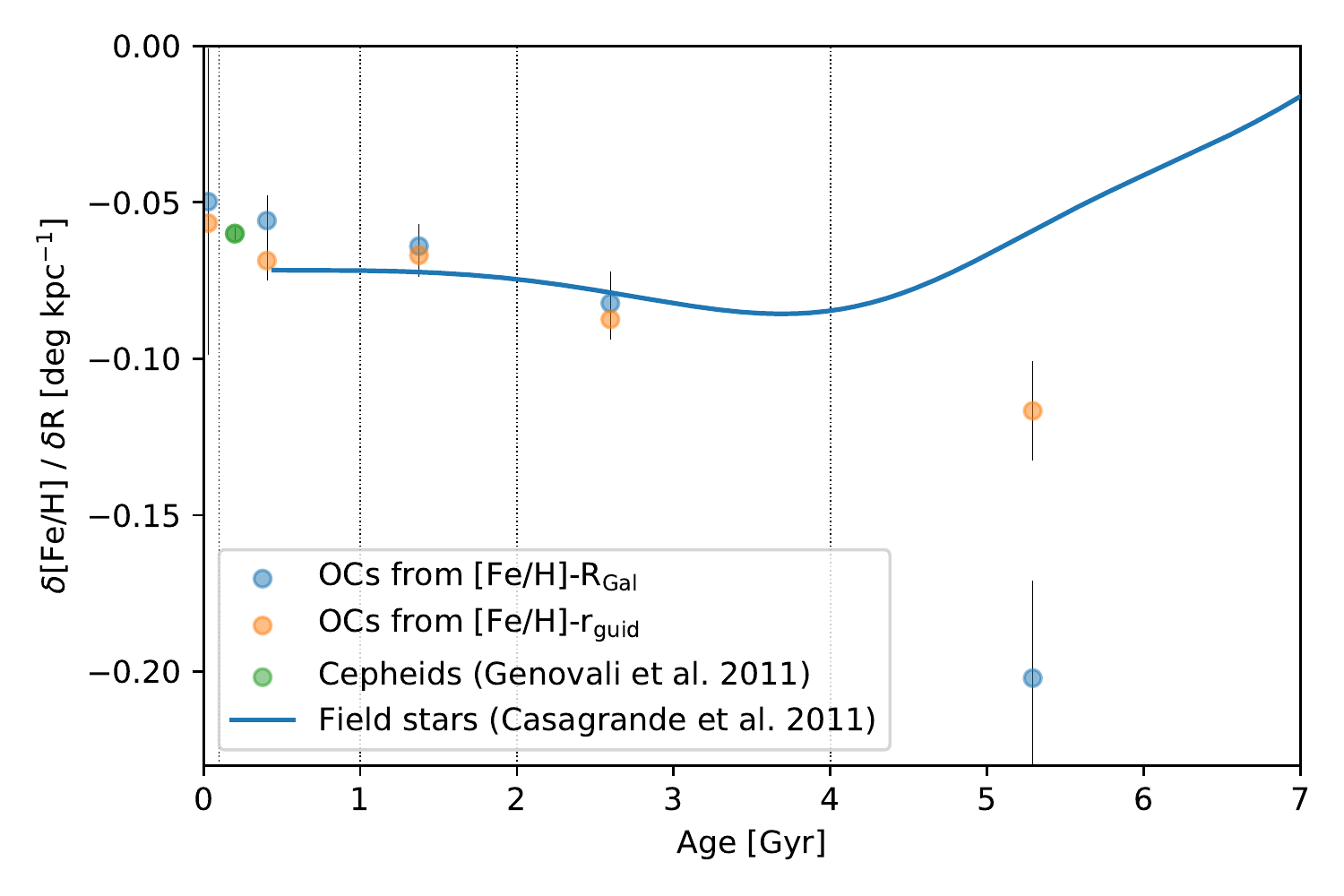}
 \caption{Age dependence of the Galactic metallicity gradient traced by open clusters in the [Fe/H]-R$_{\rm Gal}$ (blue dots) and [Fe/H]-r$_{\rm guid}$ (red dots) diagrams. The gradient age dependence traced by Cepheids (\citep{Genovali14} is
 represented by a green circle, while field stars \citep{Casagrande11} are represented by a\mbox{ solid line.}}
 \label{gradient_evolution}
\end{figure}

It is also interesting to study how the metallicity gradients evolve with time and compare them to other tracers of the metallicity distribution across the Galactic disk. In Figure~\ref{gradient_evolution}, we show the inner metallicity gradients calculated for different age bins with an orthogonal distance regression in both the [Fe/H]-R$_{\rm Gal}$ (blue circles) and [Fe/H]-r$_{\rm guid}$ (red circles) diagrams. We also compare these values with the gradient seen for Cepheids (green circle; \citep{Genovali14}), which are young stars with a range in ages that is quite limited (\mbox{$\sim$20--400 Myr}), and the gradient-age relation observed by \citet{Casagrande11} on field stars (solid line). A similar figure showing flatter gradients for younger populations, including OB stars, was previously presented in \citet{Daflon04a} for oxygen, keeping in mind that the latter element is measured in H II regions, OB stars, planetary nebulae and Cepheids.) As discussed previously in the literature \citep{Anders17,Spina17,Donor20,Spina21}, the young [Fe/H]-R$_{\rm Gal}$ gradient is flatter than the old one. Interestingly, the gradients traced by the youngest clusters are remarkably similar to those traced by Cepheids and young field stars. Nevertheless, as we move towards older bins, clusters and field stars show opposite behaviours: [Fe/H]-R$_{\rm Gal}$ gradient traced by clusters steepens, while that of field stars become flatter. Only this second behaviour is what it is expected from chemo-dynamical models of the Galactic \mbox{disk \citep{Minchev18}}. This dichotomy between clusters and field stars is still a topic of debate; however, it is likely the result of a bias that is intrinsic in the populations of open clusters and that is operated by the Galaxy itself that quickly dissipates clusters living most of their lives near the Galactic centre (for more details, see \citet{Spina21}). As it is discussed in Section~\ref{dissipation}, we require further studies of these selection effects in order to better understand how the Galaxy is shaping the current demography of open clusters. 

As it was also noticed by \citet{Spina21}, the gradient in the [Fe/H]-r$_{\rm guid}$ space has a very little evolution with time. This is probably due to the fact that action integrals are more fundamental quantities than R$_{\rm Gal}$. Differences between the distribution of open clusters within the [Fe/H]-R$_{\rm Gal}$ and [Fe/H]-r$_{\rm guid}$ diagrams need to be further investigated with larger samples and more precise astrometric data.

Finally, it is interesting to notice that, although the innermost open clusters in our sample reach [Fe/H] abundances of $\sim$0.3 dex, the young stellar populations living in the Galactic centre are roughly solar \citep{Cunha07,Davies09}. Reconciling these findings is not straightforward. Does the inner metallicity gradient follow the steep gradient that we find in the regions between say 5 and 12 Kpc, or does it rather flatten and then invert its slope as we move towards the inner regions of the Galaxy?



%

\section{Present and Future Challenges}
\label{challenges}
In recent years, the study of Galactic open clusters has been undergoing an epochal revolution due to the \textit{tsunami} of data produced by large missions and surveys such as Gaia, APOGEE, Gaia-ESO, GALAH, and many others. This unprecedented wealth of resources is pushing the field forward at high speed. Nevertheless, there is still much to understand around the chemical distribution of elements traced by open clusters. In this section, we discuss a few practical challenges that we hope would have the highest priority in order to advance the field significantly beyond its current the state-of-the-art.

\subsection{The Selection Bias}
\label{selection}

To which extent is the current census of open clusters representative of the entire population living in our Galaxy? Is there any selection bias that is preventing us from using open clusters as effective tracers of the chemical evolution of the Galactic disk?

\begin{figure}[H]

 \includegraphics[width=13.5cm]{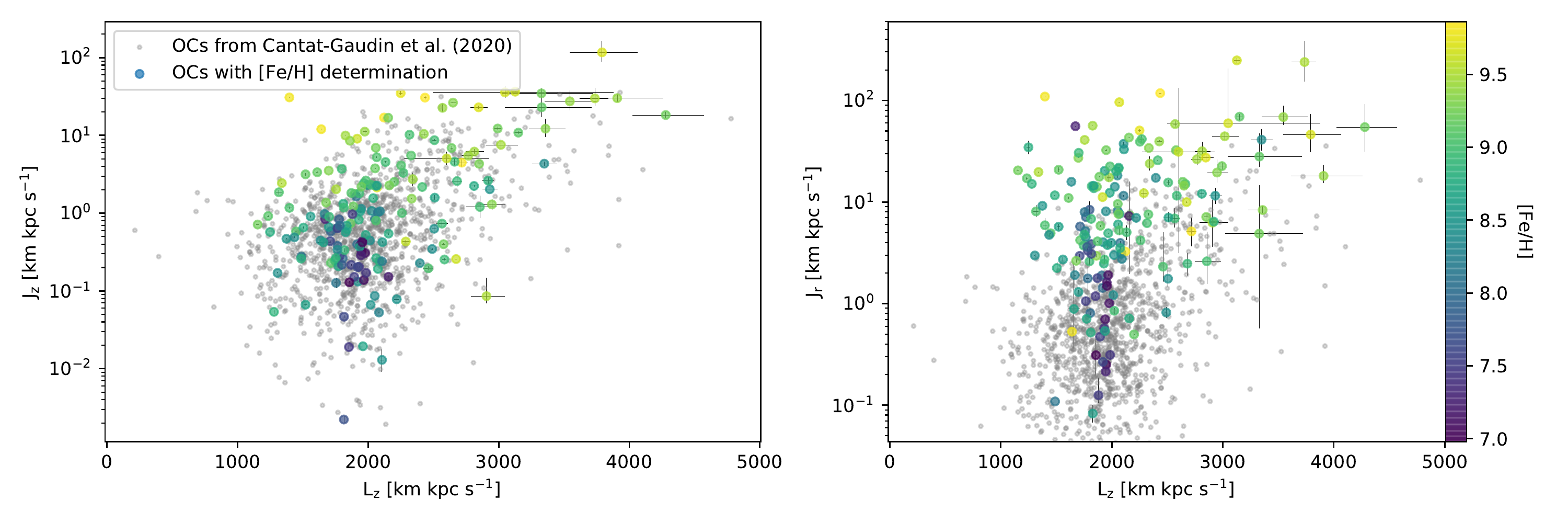}
 \caption{(\textbf{Left}) Open clusters in the action space J$_{\rm z}$-L$_{\rm z}$. The open clusters with [Fe/H] determinations that we have used to outline the metallicity gradients in Figure~\ref{gradients} are shown as coloured circles. The different colours are representative of clusters ages. The small grey dots represent the other open clusters from the census of \citet{Cantat-Gaudin20}. Among all the clusters included in that catalog, we can show only those with a radial velocity determination. The RVs of clusters without [Fe/H] determinations are taken from \citet{Tarricq21}. For these clusters, we derive the three actions L$_{\rm z}$, J$_{\rm z}$, and J$_{\rm r}$ using the same method described in Section~\ref{orbital_parameters}. (\textbf{Right}) Open clusters in the action \mbox{space J$_{\rm r}$-L$_{\rm z}$.} }
 \label{selection_bias}
\end{figure}

In Figure~\ref{selection_bias}, we show the distribution of open clusters with [Fe/H] determination (coloured circles) and that we used to outline the Galactic metallicity gradient discussed in Section~\ref{gradients}. It is readily evident that the open clusters in the outer disk---those with large L$_{\rm z}$ values---are on average 
older and also have larger values of J$_{\rm z}$ and J$_{\rm r}$ compared to the others. Therefore, the open clusters that we are using to trace the outer gradient are those that formed a long time ago that are living most of their time at large heights from the midplane, and on very eccentric orbits. Is this the effect of a selection bias due to the fact that it is \textit{easier} to collect spectra from distant stars located above the midplane? How is this potential bias affecting the metallicity gradients shown in Figure~\ref{gradients}?

An answer and a solution to this issue can be achieved by targeting open clusters that are strategically located in undersampled regions of the J$_{\rm z}$-L$_{\rm z}$ and J$_{\rm r}$-L$_{\rm z}$ diagrams. However, when we look at the distribution of open clusters that still do not have a [Fe/H] determination (grey circles), we find that very few of them are located in the outer disk and have low J$_{\rm z}$ and J$_{\rm r}$ values. Is this a real feature of the demographics 
of Galactic open clusters or instead is it the consequence of another selection bias affecting searches of open clusters? In fact, it is also in this case much \textit{easier} to spot clusters located well above the Galactic midplane, rather than those living closer to the disk where the stellar density and 
extinction are higher.

Similarly, there are no clusters with [Fe/H] determination at L$_{\rm z}$ $<$ 1100 km~kpc~s$^{-1}$, but there are also very few known open clusters below that threshold. Is it because the clusters born in the inner Galaxy are immediately dissipated due to interactions with the bar, spirals, and giant molecular clouds or instead it is that we are not seeing them because of the high stellar density and extinction?


It is well known that past and recent searches of open clusters have been strongly biased by humans decisions. Astronomers have decided the regions of the Galaxy that deserved a deeper search, the observational strategy, and the techniques of analysis, which were fine-tuned to extract clusters with specific stellar densities. Although these decisions were necessary due to the lack of resources and observing time, they have strongly biased our current catalog of clusters in a very complex way. How this selection bias is affecting what we have learnt from open clusters around the Galactic chemical evolution is still undetermined. That represents a fundamental issue that needs to be addressed. 

\subsection{The Cluster Dissipation Bias}
\label{dissipation}
Do open clusters and field stars trace the chemical map of the Galactic disk in a \mbox{similar way?}

The results presented in Figure~\ref{gradient_evolution} indicate that there may be differences between the open clusters and the field stars.
It is well known that several mechanisms can modify the orbits of stars, hence shaping their demographics across the Galaxy. For instance, stars can gain or lose angular momentum from interactions with gravitational potential (e.g., spirals, bars, molecular clouds). As a consequence, many stars have travelled a long way from the orbits in which they were born, and it is now difficult to 
ascertain where they originated. 
The situation that could be different for open clusters. In fact, the Galactic potentials that can modify stellar orbits are the same that can drive a quick disruption of open clusters. Therefore, it is possible that the open clusters that we observe today are either those that are young
enough to not have undergone numerous interactions with the gravitational potential or old clusters living most of their time far from these potentials (e.g., far from the mid plane or in the outer disk). Either way, these are the clusters that have probably conserved most of their angular momentum. 

In a recent work, \citet{Spina21} have discussed multiple hints of this dichotomy between field stars and open clusters. For instance, they have shown that old (age $>$ 2 Gyr) clusters in our census are living far from the Galactic midplane, unlike field stars that are always more densely concentrated near the midplane. Is it because the open clusters born on orbits leaning on the Galactic disk have been quickly dissipated? Or instead are we missing the old clusters on the midplane because of a selection bias?

Similarly, the metal rich clusters formed near the inner Galaxy will likely face a rapid disruption if they do not migrate outward, where the galactic potentials are weaker. This is probably how the old, metal-rich NGC~6791 survived till today. Thus, if a metal-rich open cluster that formed in the inner disk can survive only if it migrates outward, one would expect to observe the older clusters tracing a steeper gradient than that seen for young clusters and field stars. This is exactly what we observe in Figure~\ref{gradient_evolution}.


These are just realistic hypotheses, however. 
Solid evidence that opens clusters are not redistributed across the Galactic disk like field stars would represent a breakthrough in our understanding of the mechanisms that are responsible for stellar migration and cluster disruption. It would also be crucial to interpret the chemical distribution of elements traced by open clusters analogous to that of other tracers. Finally, the possibility that an existing cluster is less likely to have migrated than a coeval star in the field would justify the use of clusters as the best model-independent tracers of the role of radial migration in the Galactic disk. Therefore, a realistic comparison between the demographics of open clusters and field stars is urgently needed.

\subsection{The Role of Resonances in the Galactic Disk}
\label{resonances}

The dynamics of stars and the flows of gas across the Galactic disk obey a number of rules dictated by resonances between the frequencies of azimuthal and radial oscillations ($\Omega_{\Phi}$ and $\Omega_{\rm R}$, respectively), which are the frequency of oscillation of a star around its non-perturbed circular orbit, and the frequency of rotation $\Omega$ around the Galactic center:
\begin{equation}
    {\rm m}(\Omega - \Omega_{\Phi}) - {\rm l}\Omega_{\rm l} = 0,
\end{equation}
where m and l are small integers with m $>$ 0. All these resonances can induce overdensities or underdensities of stars and gas within the Galactic disk. They can even act as barriers preventing matter from freely flowing across the disk.

It is expected that this complex dynamical substructure of the Galactic disk would produce visible signatures in the chemical distribution of elements \citep{Wheeler21}. Therefore, features may be noticeable in the metallicity gradient outlined by open clusters at the corotation (l,m = 0), or at the Lindblad Resonances (l = $\pm$1, m = 2). A possible consequence of these resonances could be the crest visible at L$_{\rm z}\sim$2000--2200 km~kpc~s$^{-1}$ from the LOWESS in Figure~\ref{gradients} bottom. Furthermore, the break between the inner and outer gradients and the flatness of this latter also deserve an explanation within this context \citep{Lepine11}.

Now that Gaia data are unveiling interesting dynamical patterns within the Galactic disk \citep{Trick21}, we believe that the effect of these resonances 
deserves further investigations in relation to the chemical distribution of elements traced by open clusters.

\subsection{The Role of the Galactic Warp}
\label{warp}

\begin{figure}[H]

 \includegraphics[width=13.5cm]{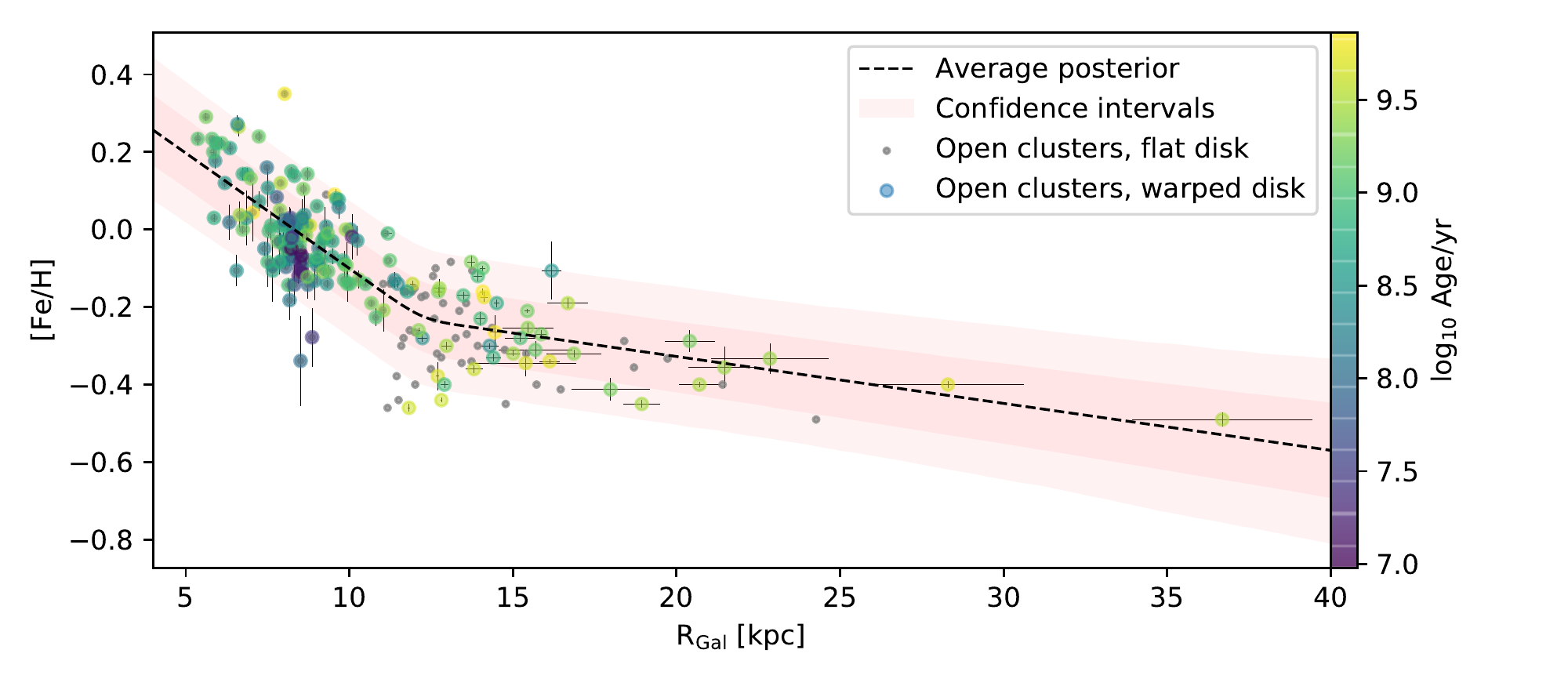}
 \caption{[Fe/H] values determined for open clusters as a function of their Galactocentric distances R$_{\rm Gal}$. Grey dots represent the open clusters under the \textit{flat disk} assumption, while the coloured circles are the clusters after the correction for the Galactic warp. The latter are colour coded as a function of their age. Red shaded areas represent the 68 and 95$\%$ confidence intervals of the models resulting from the Bayesian inference, while the black dashed
line traces the most probable model.}
 \label{gradients_warp}
\end{figure}

Our Galaxy, like many large spiral galaxies, is warped and flared in its outskirts. 
The Galactic warp and flare are traced by various populations: gas such as atomic H{\sc i} e.g., \citep{Nakanishi2003, levine06, koo17}, ionized hydrogen e.g., \citep{cersosimo09} and molecular clouds e.g., \citep{may97, nakagawa05, xu18}, 
and stellar populations e.g., \citep{hammersley11, poggio17, li19, Chrob2020}, including Cepheids e.g., \citep{feast14,  Chen2019, skowron19} and open clusters e.g., \citep{carraro16, Cantat-Gaudin20}. 
\citet{amores17} investigated the dependence with the age of the structural parameters of the outer disc, including warp, flare, and disc truncation, finding strong evidence that the thin disc scale length, as well as the warp and flare shapes, changes with time. 
As described in \citet{amores17}, this might be due to a misalignment of the disc inside the dark halo, which can change with time, provoking a precession, or to the interaction of our Galaxy with the Magellanic Clouds. 
If the warp has a dynamical origin, tracers of different ages might show a different behaviour \citep[see also][]{lopezcorredoira07}. 
Using the results for the time-dependence of the warp in our Galaxy from \citet{amores17}, we have investigated the effect of the warp on the shape of the radial metallicity gradient. 
Since our sample of clusters spans a large range in ages, we have corrected their R$_{\rm GC}$, de-projecting it along the warped disc as in \citet{amores17}, considering the variation with time of both the amplitude, the starting radius and the angle of the warp.  
The effect is indeed negligible for the youngest clusters, while it affects mainly the older clusters (age $\gtrsim$ 1 Gyr), located in the outer disk (R$_{\rm Gal}$ $\gtrsim$ 10 kpc). 
The results are shown in Figure~\ref{gradients_warp}, where the grey dots represent the open clusters under the \textit{flat disk} assumption, while the coloured circles are the clusters after the correction for the Galactic warp.

Using the R$_{\rm Gal}$ corrected for the Galactic warp, we repeat the analysis described in Section~\ref{gradients}. The metallicity gradient of the warped disk is plotted in Figure~\ref{gradients_warp} and the numerical results are listed in Table~\ref{posteriors_tab} bottom panel. Since the open clusters in the warped disk are located at greater distances, the new slope is slightly flatter than that found through the \textit{classic} analysis. As a consequence, the break between the inner and outer disks also appears to be more distant from the Galactic centre. However, all the differences the gradient parameters are within their uncertainties. Nevertheless, we also note that the posteriors' confidence intervals are tighter for the \textit{warped disk} than those found for the \textit{flat disk}. Finally, it is remarkable to see how the correction for the Galactic warp has moved the two outermost clusters of more than 5 kpc away from the Galactic centre.

The analysis described in this section is a simple illustration 
highlighting how the Galactic warp affects the radial metallicity gradient that we observe today. It is well known that the warp shows a complex structure, and it is strongly asymmetrical. There is no doubt that forthcoming Gaia data releases will pave the way for more accurate studies of the Galactic warp. These studies will be fundamental to trace a more realistic Galactic metallicity gradients.

\subsection{Intra- and Inter-Clusters Chemical Homogeneity}
\label{chem_tagging}

Regardless of the precision achievable in elemental abundances, the success of chemical tagging relies on the significance of two other 
critical factors: (i) the level of chemical homogeneity among stellar members of the same cluster; (ii) the chemical diversity among open clusters.

These factors can be directly measured using distance metrics similar to that employed by \citet{Mitschang13}:

\begin{equation}
    {\rm D_{C} = \sum^{N_{C}}_{C}}\omega_{\rm C}{\rm \frac{|A^{i}_{C}-A^{j}_{C}|}{N_{C}}},
\end{equation}
where C is a defined chemical space formed by N$_{C}$ elements with A$_{C}$ abundances. A weight $\omega_{C}$ is assigned to each element. This Manhattan-like distance is much less affected by outliers than the classical Euclidean distance and, for that reason, it should be preferred in studies of chemical tagging, especially those based on large volumes of data produced by surveys. \citet{Mitschang13} used the score described above to probe the intra- and inter-cluster level of homogeneity. The intra-cluster homogeneity level is given by the typical D$_{C}$ distribution calculated among stars belonging to the same association, while, for the inter-cluster homogeneity level, D$_{C}$ is computed from stars that are not members of the same cluster. This experiment has shown in practice that open clusters are ideal empirical calibrators of chemical tagging techniques. In fact, through them, we can understand chemical tagging in a practical sense, and probe different techniques of analysis (see \mbox{also \citep{Blanco-Cuaresma18}}). 

It should be stressed, however, that intra- and inter-cluster homogeneity levels could be non-universal factors. For instance, old massive open clusters may be formed by multiple generations of stars and---similarly to what it is observed in globular clusters---these generations may be distinguishable also in the chemical space. Nevertheless, to date, all open clusters have been shown to be composed by single stellar populations \citep{Bragaglia12,Carraro14}. A previous claim of chemical inhomogeneity for NGC~6791 \citep{Geisler12} has not been confirmed \mbox{in \citep{Bragaglia14,Cunha15}} and more recently in \citep{Villanova18}.

Other sources of chemical inhomogeneities have been found in open clusters, such as those related to atomic diffusion \citep{Souto18,BertelliMotta18} or planet engulfment events \citep{Spina15,Spina18,dorazi20}. The first can be observed by comparing the chemical patterns of main-sequence stars to those of giant stars (see Section~\ref{atomicdiffusion}) or fully convective M dwarfs \citep{Souto21}. Instead, the chemical signatures of planet engulfment events are mostly limited to G and late F-type stars, which are characterised by extremely thin convective zones that can be easily polluted by accreted material \citep{Spina21b}.

It is also very likely that the inter-cluster homogeneity level may be a function of the Galactocentric radius. The inner Galactic disk has evolved much faster than the outer disk. Therefore, the open clusters closer to the Galactic centre may be more chemically diverse than those formed at the outskirt of the Galaxy \citep{casali20}. This would imply that chemical tagging in the inner Galaxy could be more efficient than in other regions of the disk.

In conclusion, despite the important works and attempts that have been conducted in the field, there is still much to understand around the intra- and inter-cluster homogeneity levels. Open clusters (and wide binary pairs) should be regarded at the centre of future studies in this field.

\subsection{Spectroscopic Analysis of Young Stars}
\label{young_stars}

\begin{figure}[H]

 \includegraphics[width=13cm]{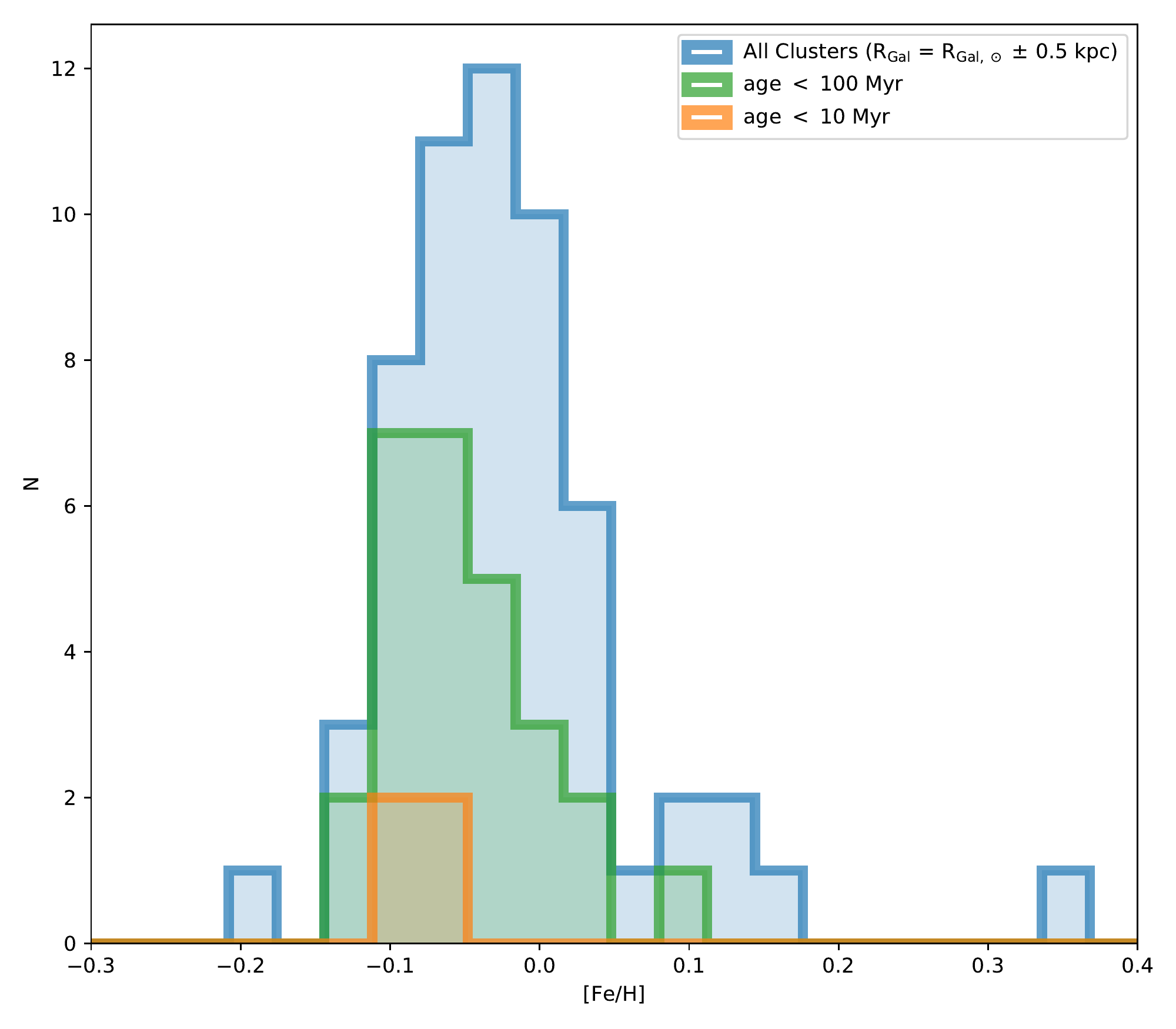}
 \caption{Metallicity distributions of open clusters from the \textit{gold} sample located at R$_{\rm Gal}$ within R$_{\text{Gal},}\pm$0.5~kpc.}
 \label{hist_age}
\end{figure}

 Figure~\ref{hist_age} shows the metallicity distribution of the open clusters from the \textit{gold} sample located at R$_{\rm Gal}$ within R$_{\text{Gal},}\pm$0.5~kpc. While these clusters span a wide range of [Fe/H] values between $-$0.2 and +0.35 dex, the fraction of clusters younger than 100~Myr is mostly restricted to sub-solar metallicities. Furthermore, all the youngest associations in this sample (age $<$ 10 Myr) have on average even lower metal content. 
 
This evidence is clearly at odds with chemical evolution models of our Galaxy \citep{Magrini09,Minchev13,Minchev14}, which predict an increase of the metallicity with time. 
Such contention between theory and observations was previously noticed both within the solar vicinity \citep{James06,DOrazi09,DOrazi11,Biazzo11a,Biazzo11b,Spina14,Spina21} and beyond \citep{Spina17}.

Recent works have shown that the anomalously low metal content of the youngest stars in our Galaxy is not the result of chemodynamical processes acting within the disk. Instead, they are caused by a synthetic gap in models of stellar atmospheres. The latter typically neglect the effects of stellar activity and magnetic fields on the formation of spectroscopic lines. These effects are particularly strong in young stars \citep{Lorenzo-Oliveira18}. As a consequence of this approximation, the stellar atmospheres of active stars appear to be more metal poor than 
they actually are \citep{Galarza19,Baratella20,Spina20}.


This poses a challenge for stellar spectroscopy. Most of the open clusters in our Galaxy dissolve on timescales that are typically shorter than a few Gyr due to the interactions with the bar, spirals and giant molecular clouds \citep{Anders21}. This implies that stellar members of these clusters are in large part young and active stars. The lack of a reliable methodology to derive their 
chemical content 
ultimately prevents us 
from using open clusters as effective tracers of the Galactic chemical evolution.

Important efforts have been done in developing new approaches of spectroscopic analysis based on the use of ad hoc linelists composed exclusively by those absorption features that are less affected by chromosperic activity \citep{Baratella20}. Interestingly, these works have shown that the youngest clusters in the solar vicinity actually have super-solar metallicities ranging between between 0.04 and 0.12 dex. Although this work represents an important step towards the solution of the problem, developing new models of stellar atmospheres and new techniques of spectroscopic analysis that could fully take into account the effect of stellar activity would represent the optimal answer to this critical issue.

\subsection{Analysis of Cool Stars}
\label{coolstars}

\begin{figure}[H]

 \includegraphics[width=13.5cm]{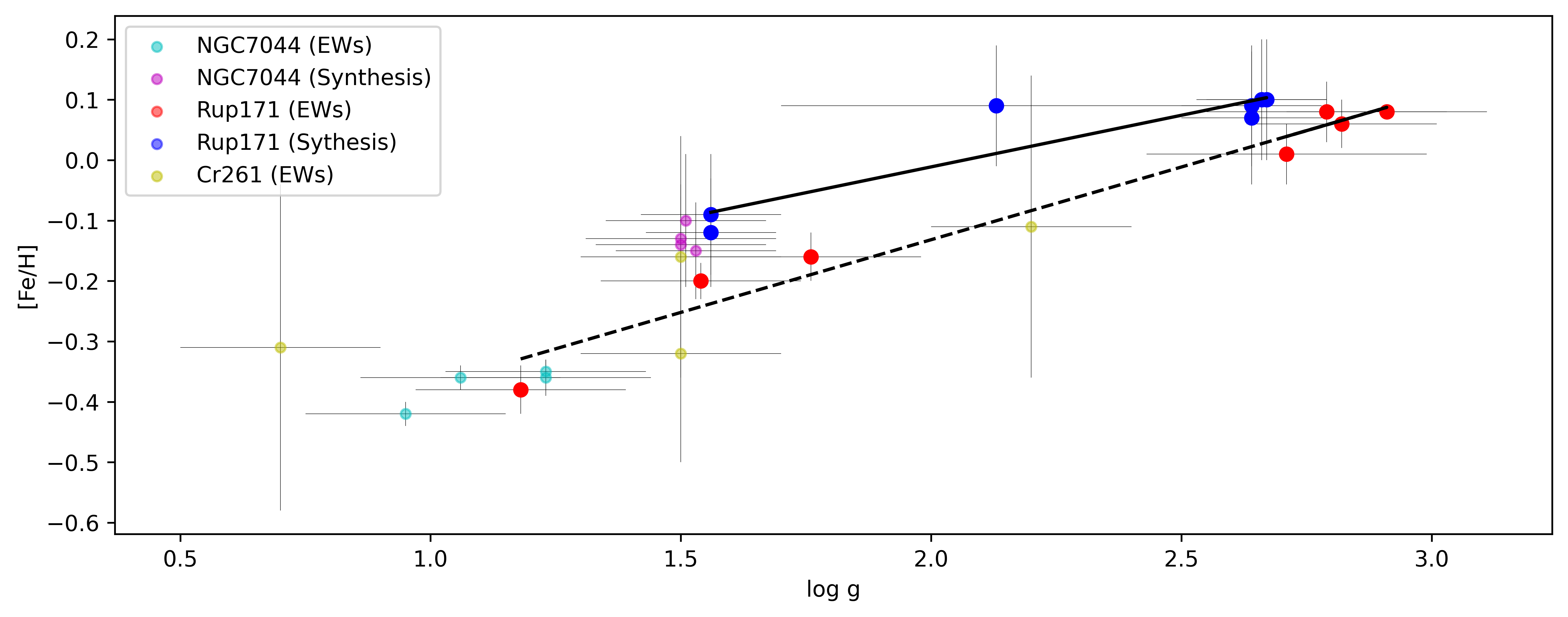}
 \caption{The figure shows [Fe/H] as a function of log~g in a sample of member stars in three open clusters, in which cool giants were observed: NGC~7044 and Rup171 from \cite{casali20} analysed with both the EWs and the spectral synthesis, and Collinder~261 (only with EWs) from \cite{friel03}. Colours and symbols are described in the legend.The continuous line is the linear regression to the results for Rup~171 obtained with spectral synthesis, while the dashed-line is the linear regression of the results from EWs for Rup~171.}
 \label{fig:cool}
\end{figure}

In some distant or particularly extinct clusters, only the brightest and coolest giants have been targeted.
The spectra of cool stars, with T$_{\rm eff}\leq$ 4300~K and log~g $\leq$ 1.8, are usually more difficult to analyse, in particular their surface gravity, and, consequently, their T$_{\rm eff}$ and metallicity. 
 The differences between the abundances from hotter and cooler stars in the same cluster, especially when their spectra are analysed through EWs, can be related, e.g.,  the failure of model atmospheres at low effective temperatures see, e.g., \citep{bessel98, short03} or to the definition of the continuum near to the lines of interest in spectra dominated by line crowding (i.e., in particular high-metallicity giant stars) e.g., \citep{casali20}. On the other hand, spectral synthesis is less prone to continuum setting and blending effects, producing more solid determination of the stellar parameters also in cool giant stars. 
In Figure~\ref{fig:cool}, we show [Fe/H] as a function of log~g in a sample of member stars in three open clusters, in which cool giants where observed: NGC~7044 and Rup171 from \cite{casali20}, analysed with both the EWs and the spectral synthesis, and Collinder~261 (only with EWs) from \cite{friel03}. There is a clear decreasing trend of [Fe/H] with decreasing log~g, much more pronounced in the analysis with EWs.  The metallicity of NGC7044, in which only cool red giants are observed, is lower than we might expect for a cluster of $\sim$2 Gyr located in the solar neighbourhood. 
Considering only the stars of Rup~171, which are the most numerous in the sample shown in the figure and which cover the largest log~g interval, we can compare the trend obtained with the spectral synthesis and that obtained with the EWs. Although there is an improvement in the analysis obtained with spectral synthesis (low gravity stars have metallicity closer to high gravity stars), the trend remains, with a slope of \mbox{0.17 $\pm$ 0.03} to be compared with a slope of \mbox{0.24 $\pm$ 0.02} for the analysis obtained \mbox{with EWs.} 
Therefore, it is important to take this potential problem into account when dealing with clusters where only cool giant stars are observed, even if they are analysed with spectral synthesis, we might still have non-negligible effects on the derived parameters, particularly on metallicity. We suggest excluding the abundances in cool giants in the calculation of average cluster abundances.


\subsection{Atomic Diffusion}
\label{atomicdiffusion}
Despite the problems highlighted in Section~\ref{coolstars}, evolved stars will certainly play a central role in tracing the chemical distribution of elements with open clusters. The chemical abundances that we measure in the stellar atmosphere can change along the different different evolutionary phases. This chemical abundance variation results from diffusion processes have been probed both in the metal-poor globular clusters \citep{Korn07,Lind08,Nordlander12}, and in open clusters \citep{Souto18,BertelliMotta18}. 

Atomic diffusion changes the surface chemical composition of stars during their main-sequence phase mainly because of gravitational settling that induces different elements to sink towards the interior of the star. The magnitude of this effect reaches its maximum at the turn-off, where the stellar atmosphere can be up to 0.10 dex poorer in Fe than its pristine composition \citep{Dotter17,Souto18}. After the turn-off, the outer convective zone becomes deeper and material from the stellar interior is brought back to the surface: the stellar atmosphere reacquires its initial chemical pattern.

The chemical composition changing along the evolutionary phases of stars can introduce small biases when we want to trace the chemical distribution of elements across the Galaxy. In fact, spectroscopic surveys typically observe a mixture of giants and dwarfs stars, with the evolved stars being generally observed at large distances, while dwarfs are often used as metallicity tracers around the solar location. A counter example is represented by APOGEE, which mostly observes evolved stars. 


Future works should take this bias into account when significant. This is especially important for those interested in measuring the chemical scatter between open clusters at each Galactocentric radius, which is often interpreted as an indirect quantification of cluster migration e.g., \citep {Quillen18a,Spina21}.

\section{The Radial Gradient of the Milky Way in the Extragalactic Framework}

\textls[-25]{The Milky Way is a benchmark to study and understand the family of disk \mbox{galaxies \citep{BlandHawthornGerhard17}.}} 
Only for the MW can we access the full star formation history of a galaxy, observing from the faint ancient dwarf stars to the young massive supergiant ones. The MW is one of the myriad of spiral galaxies of the Universe. However, Galactic studies will continue to play a fundamental role far into the future because there are measurements that can only be made in the near field. 
The MW is a luminous barred spiral with a central box/peanut bulge, dominated by its disk, and with a
diffuse stellar halo. 
In a way, the Galaxy is a rather common spiral galaxy, located in a low density environment with a typical star formation rate, baryon fraction and stellar mass \citep{kormedy10}. On the other hand, some of its characteristics are quite unusual: 
following \citet{BlandHawthornGerhard17}, the MW falls in the sparsely
populated “green valley” region of the galaxy colour-magnitude diagram, in transition between the ‘red sequence’ of galaxies and the ‘blue cloud’
\citep{MutchCrotonPoole11}.  In addition, the presence of two luminous dwarf galaxies (the Small and Large Magellanic Clouds, SMC and LMC) orbiting around the MW are very uncommon \citep{robotham12}. 

\begin{figure}[H]

 \includegraphics[width=13.5cm]{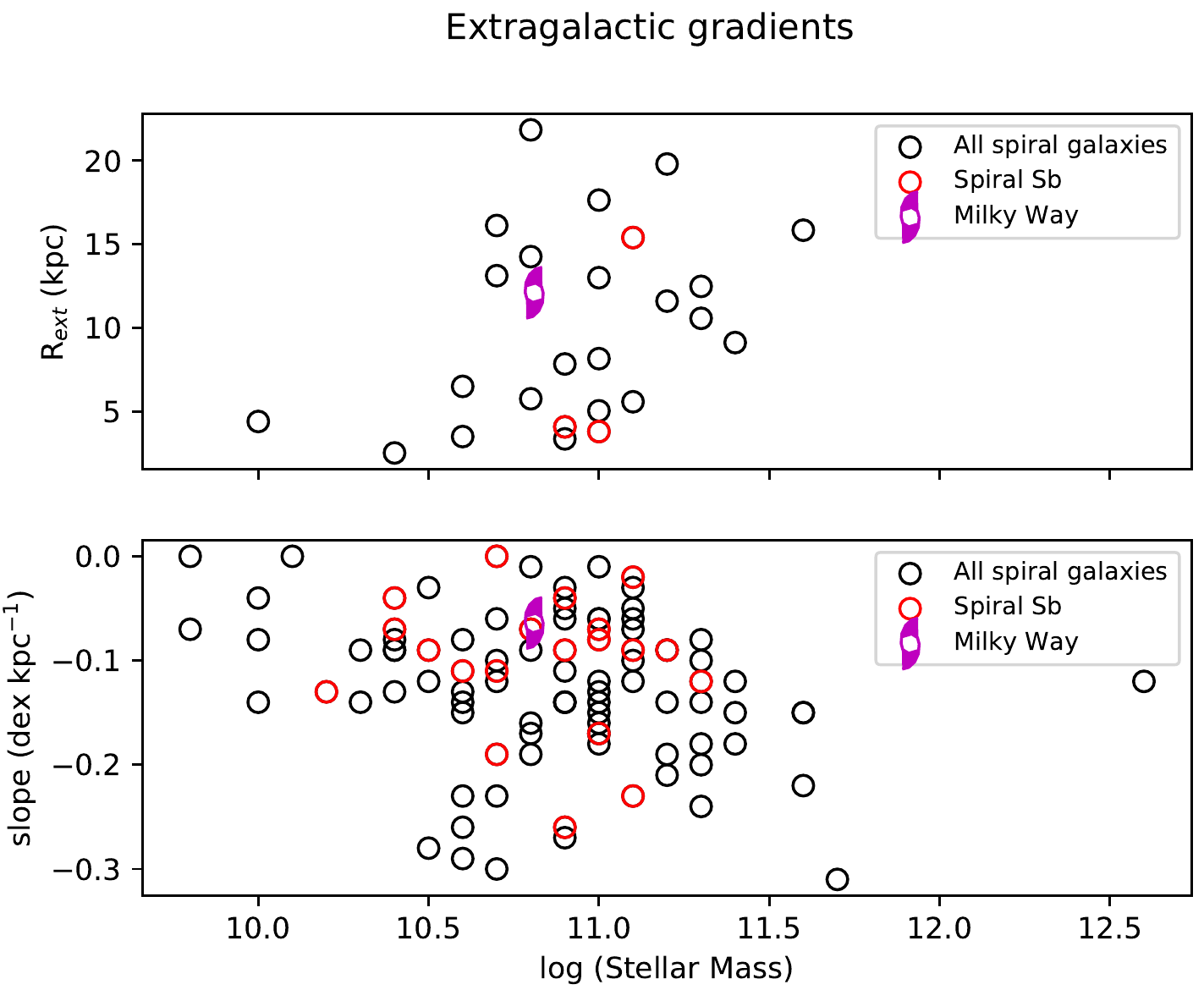}
 \caption{The figure shows the R$_{\rm ext}$, the galactocentric radius at which the change of slope occurs, and the slope of the gradient in a sample of 95 spiral galaxies observed with MUSE@VLT in \citet{sanchez18}. Colours and symbols are described in the legend.}
 \label{fig:extra}
\end{figure}

In the framework of the surveys aimed at studying the properties of nearby galaxies, it is interesting to compare the shape of the Galactic radial metallicity gradient with those of possible morphological analogues, keeping in mind that no two galaxies are exactly identical even if their morphology is very similar see, e.g., \citep{sanchez18,bresolin19,boardman21, zurita21a, zurita21b}. 
This comparison allows us to highlight common features between galaxies of similar morphological types, and at the same time to see which features are unique to our Galaxy. 
In this review, we compared the Galactic gradient slope and galactocentric radius at which the change of slope occurs, with the gradients observed in a sample of 95 spiral galaxies observed with Multi Unit Spectroscopic Explorer MUSE, see \citep{bacon10} at VLT by \mbox{\citet{sanchez18}.} 
Both the slope and the radial position of the outer flattening in the MW, expressed in physical units, are in agreement with those observed in galaxies of similar morphological type, although there is a considerable spread, even in the sample of Sb galaxies. 
The large dispersion in the amplitude of the gradient slopes might indicate a possible dependence of the gradient with some particular property of the galaxies. In particular, \citet{sanchez18} found that the steepest gradients are related to the presence of an inner drop or of an outer flattening. They suggested that radial motions might be playing an important role shaping the abundance profiles, and they might cause the presence of these features.
The variety of behaviours at each given stellar mass might be homogenised when considering the gradients reported on the scale of the effective radius \citep{sanchez14}. However, it is not trivial to have an estimate of the Galactic effective radius; in previous works, the MW disc scale length has been estimated to be atypically short, 2.15~kpc, see \mbox{e.g., \citep{Bovy13, Licquia16}}, much shorter than those in typical MW analogue galaxies. 
\citet{boardman20b} compared the gradient of the
Milky Way with a sample of MW-analogue galaxies in the Mapping Nearby Galaxies at APO (MaNGA) sample. With their definition of MW analogues, based on the bulge-to-total ratio and not only on morphology, as opposed to what is shown in our Figure~\ref{fig:extra}, where only morphology was taken into account, they found that 
the Galactic gradients are steeper (in dex kpc$^{-1}$) than for a typical MW-analogue of their sample (see their Figure 10).
 Part of the discrepancy between the MW and its analogues might be related to the difference between the internal perspective offered by the MW and the external perspective available for other
galaxies, and to our limited estimation of its scale-length, which is carried out via star-count analysis and differs significantly from the photometric methods employed in other galaxies \citep{boardman20a}. 
In particular, as pointed out by \mbox{\citet{boardman20b}}, we should be careful to compare ``apple with apple'' in the relative comparison between the MW and other galaxies, taking into account the differences due to integrated light of external galaxies versus the  measurements of individual objects in the MW; the binning applied to the IFU data which might flatten the measured stellar population ratios see, e.g., \citep{Ibarra19}; the differences in gradient between mono-age and multi-age populations; see, e.g., \citep{Minchev13, Minchev14}.

The next few years will allow us to characterise our Galaxy even better.
Thanks to the {\em Gaia} satellite data, we are having dynamical models of the MW, allowing us to measure the mass distribution in the disk, and its scale lengths; see, e.g., \citep{Nitschai20}. 
The new spectroscopic surveys with WEAVE \citep{Dalton16}, 4MOST \citep{DeJong12}, and MOONS \citep{Cirasuolo14}, SDSS V MWM (using the APOGEE spectropgraphs) will enlarge the number of spectroscopic observations in open clusters, which remain among the best tracers of the Galactic metallicity gradient. At the same time, observations with {\em Gaia} and the Rubin telescope \citep{Ivezic19, Prisinzano18} will lead to the discovery of new clusters, complementing our view and overcoming some observational biases that affect our current knowledge of the Galactic cluster population. 
With such in-depth knowledge of the MW, we will be able to compare it in detail against 
its analogues, relating eventual differences  with physical reasons, e.g., the unusual compactness of the MW for a galaxy of its type, its low density environment, or the interaction with Magellanic Clouds, and locating it in the framework of a wider galaxy population.

\section{Summary}
In this paper, we make use of data collected for open cluster members by high-resolution spectroscopic surveys and programmes: APOGEE, Gaia-ESO, GALAH, OCCASO, and SPA. These data have been homogenised and then analysed as a whole. The resulting catalogue contains [Fe/H] and orbital parameters for 251 Galactic open clusters. 

The slope of the radial metallicity gradient obtained from 175 open clusters with high-quality metallicity determinations is $-$0.064 $\pm$ 0.007 dex kpc$^{-1}$ (see Figure~\ref{gradients} top, Figure \ref{posteriors} left, Table~\ref{posteriors_tab}). The radial metallicity distribution traced by open clusters flattens beyond R$_{\rm Gal}$12.1 $\pm$ 1.1 kpc. 

We also investigate the distribution of open clusters in the [Fe/H]-L$_{\rm z}$ diagram (see Figure~\ref{gradients} bottom, Figure \ref{posteriors} right, Table~\ref{posteriors_tab}). The slope we obtain is $-$0.31 $\pm$ 0.02 10$^{3}$ dex km$^{-1}$ kpc$^{-1}$ s. The distribution flattens beyond L$_{\rm z}$ = 2769 $\pm$ 177 km kpc s$^{-1}$. 

We notice that the metallicity scatter in the [Fe/H]-L$_{\rm z}$ diagram is lower than that obtained from R$_{\rm Gal}$ (Figure \ref{posteriors}, Table~\ref{posteriors_tab}). Furthermore, the  [Fe/H]-L$_{\rm z}$ slope appears to be more stable with time than the classical radial metallicity slope (see Figure~\ref{gradient_evolution}). These findings suggest that L$_{\rm z}$ is a better suited quantity than R$_{\rm Gal}$ to characterise the metal content across the Galaxy.

In Section~\ref{challenges}, we review some high-priority practical challenges around the study of open clusters that will significantly push our understanding beyond the state-of-the-art. Namely, our knowledge on the open cluster demography is affected by selection biases which are preventing us from performing statistical studies of their populations and using them as effective tracers of the Galactic evolution (see Section~\ref{selection}). It is very likely that the Galaxy is also applying a selection of open clusters, dissipating those that are living most of their lives close to the midplane and the inner disk. Understanding this cluster dissipation bias is also fundamental to carry out meaningful comparison between the open clusters and field stars populations (see Section~\ref{dissipation}). In addition, we still need to understand which is the role of resonances in the Milky Way disk and of the Galactic warp in the spatial distribution of open clusters, with strong implications also on radial metallicity distribution that they trace (see Sections~\ref{resonances} and \ref{warp}). Given recent attempts for chemical tagging, it is fundamental to understand which is the intra- and iner-cluster chemical homogeneity at different locations within the Galactic disk (see Section~\ref{chem_tagging}). Finally, we still need to address how the anomalous metallicity of young stars, cool stars and atomic diffusion have affected studies of the radial metallicity distributions of open clusters (see Sections~\ref{young_stars}--\ref{atomicdiffusion}).

Finally, in Section~\ref{gradients_warp}, we compare the shape of the Galactic radial metallicity gradient to those of other spiral galaxies. The large dispersion in the amplitude of the gradient slopes that we observe for galaxies similar to our own Milky Way might indicate a possible dependence of the gradients to some particular properties of the galaxies, such as the total mass and radial motions of stars and gas.

\vspace{6pt} 



\authorcontributions{Conceptualisation L.S., L.M., K.C.;  analysis L.S.; investigation L.S., L.M., K.C.; data curation L.S.; writing L.S., L.M.; review and editing L.S., L.M., K.C. All authors have read and agreed to the published version of the manuscript.
}

\funding{L.S. is supported by the Italian Space Agency (ASI) through contract 2018-24-HH.0 to the National Institute for Astrophysics (INAF).

}

\institutionalreview{Not applicable.
}

\informedconsent{Not applicable.

}

\dataavailability{Table~\ref{dataset} is fully available at the CDS.
} 

\acknowledgments{The authors are extremely grateful to Simone Daflon for the interesting conversations about open clusters and the radial metallicity gradients.
L.S. also acknowledges Tristan Cantat-Gaudin, Neige Frankel, and Yuan-Sen Ting 
for the inspiring discussions about selection biases. L.S. is supported by the Italian Space Agency (ASI) through contract 2018-24-HH.0 to the National Institute for Astrophysics (INAF).

}

\conflictsofinterest{The authors declare no conflict of interest.
} 




\begin{adjustwidth}{-4.6cm}{0cm}			
\printendnotes[custom]			
\end{adjustwidth}



\begin{adjustwidth}{-\extralength}{0cm}
\reftitle{References}



\end{adjustwidth}

%


\end{document}